 \documentclass[3p,times]{elsarticle}
\usepackage{amssymb}
\usepackage{graphics}
\usepackage{psfrag}
\usepackage{epsfig}
\def\bea{\begin{eqnarray}}
\def\eea{\end{eqnarray}}
\def\be{\begin{equation}}
\def\ee{\end{equation}}

\newcommand{\pbarp}{{\bar p p}}
\newcommand{\nbarn}{{\bar n  n}}
\newcommand{\NbarN}{{\bar N  N}}
\newcommand{\ebare}{{e^+e^-}}

\newcommand{\lbarl}{{\bar \Lambda \Lambda}}
\newcommand{\lcbarlc}{{\bar \Lambda_c \Lambda_c}}

\newcommand{\sbarl}{{\bar \Sigma^0 \Lambda}}

\newcommand{\sbars}{{\bar \Sigma \Sigma}}
\newcommand{\spbarsp}{{\overline{\Sigma^+} \Sigma^+}}
\newcommand{\sobarso}{{\bar \Sigma^0 \Sigma^0}}

\begin{document}
\begin{frontmatter}
 \title{The electromagnetic form factors of the $\Lambda$ in the timelike region}
 \author[J]{J. Haidenbauer}
 \author[B,J]{U.-G. Mei{\ss}ner}
 \address[J]{Institute for Advanced Simulation, Institut f{\"u}r Kernphysik and
 J\"ulich Center for Hadron Physics, Forschungszentrum J{\"u}lich, D-52425 J{\"u}lich, Germany}
 \address[B]{Helmholtz Institut f\"ur Strahlen- und Kernphysik and Bethe Center
  for Theoretical Physics, Universit\"at Bonn, D-53115 Bonn, Germany}

\begin{abstract}
The reaction $e^+e^- \to \bar \Lambda \Lambda$ is investigated for energies close to the threshold. 
Specific emphasis is put on the role played by the interaction in the final 
$\bar \Lambda \Lambda$ system which is taken into account rigorously. For that interaction a variety 
of $\bar \Lambda \Lambda$ potential models is employed that have been constructed for the 
analysis of the reaction $\bar p p \to \bar \Lambda \Lambda$ in the past. 
The enhancement of the effective form factor for energies close to
the $\bar \Lambda \Lambda$ threshold, seen in pertinent experiments, is reproduced.
Predictions for the $\Lambda$ electromagnetic form factors $G_M$ and $G_E$
in the timelike region and for spin-dependent observables such as 
spin-correlation parameters are presented.

\end{abstract}

\end{frontmatter}

\section{Introduction}
\label{sec:intro}

The electromagnetic form factors (EMFF) of the nucleons in the timelike region
have been studied intensively over the last few years, see the recent review \cite{Denig}.
Information on these quantities is accessible in the $\pbarp$ annihilation
process $\pbarp \to \ebare$ \cite{Bardin}, 
and likewise in the reactions $\ebare \to \pbarp$ and $\ebare \to \nbarn$ 
from where most of the recent data emerged \cite{Aubert0,Lees} and where still 
experiments are onging \cite{Achasov:2014,Ablikim:2015,Akhmetshin:2015}.
New experiments for $\pbarp \to \ebare$ are in planning \cite{Singh:2016}.
One particular feature that attracted wide attention was the observation 
of a strong enhancement of the proton EMFFs close to the $\pbarp$ 
threshold, i.e. at momentum transfers $q^2\simeq (2 M_p)^2$.
This behavior was first detected in the PS170 experiment~\cite{Bardin},
in a measurement of $\pbarp \to \ebare$ at the CERN Low Energy Antiproton Ring (LEAR), 
and confirmed in recent years in experiments with high mass resolution by the BaBar 
collaboration for the time-reversed process $\ebare \to \pbarp$ \cite{Aubert0,Lees}.
The majority of theoretical studies 
\cite{Sibirtsev:2006,Dmitriev07,Baldini:2007,Chen:2010,Dalkarov:2009,Haidenbauer:2014,Lorenz:2015,Dmitriev:2015}
attributed this strong enhancement to effects of the $\pbarp$ 
interaction in the final state. This final state interaction (FSI) enhances the near-threshold
$\ebare \to \pbarp$ cross section as compared to the phase space and, in turn, 
yields (effective) proton EMFFs that peak at the threshold and then fall off 
rapidly with increasing energy.  
Refined calculations of the $\ebare \to \pbarp$ cross section (and/or EMFFs) -- like the
one performed by our group \cite{Haidenbauer:2014}
which relies on a formally exact treatment of the effects from the $\pbarp$ interaction 
in the final state and utilizes $\NbarN$ potentials that are constructed in the
framework of chiral effective field theory \cite{Kang:2013} and fitted to results of a 
partial wave analysis of $\pbarp\to \pbarp $ and $\pbarp\to \nbarn$ scattering 
data \cite{Zhou:2012} --
yield results that are in excellent agreement with the near-threshold data.

In the present paper we take a look at the electromagnetic form factors in the
timelike region of another baryon, namely those of the $\Lambda$ hyperon. 
In this case much less measurements
have been published \cite{Bisello:1990,Aubert:2007,Dobbs:2014,Morales:2016} 
(and with significantly lower mass resolution) and there are only 
few theoretical studies \cite{Baldini:2007,Dalkarov:2009,Faldt:2013,Faldt:2016}.
Anyway, the available data suggest that the $\Lambda$ form factor exhibits a
near-threshold behavior very much similar to that of the proton \cite{Aubert:2007}.
Therefore, it is interesting to see whether we can reproduce that property within
the same framework we have set up and employed successfully to the proton form factor 
\cite{Haidenbauer:2014}. 
Another and equally important motivation for our study are ongoing pertinent experiments 
by the BESIII collaboration at the BEPCII $\ebare$ collider in Beijing. 
In this experiment, the self-analyzing character of the weak decay $\Lambda \to p\pi^-$ 
will be exploited so that it is possible to determine also the polarization as well as 
spin-correlation parameters for the reaction $\ebare \to \lbarl$ 
\cite{Morales:2016,Tord}. Those observables allow one to determine not only the (effective) 
form factor but also the relative magnitude of the two electric form factors $G_M$ and $G_E$ 
and even the phase between them \cite{Denig,Faldt:2016}. 
Within our approach we can make predictions for those observables. 

Our calculation for $\ebare \to \lbarl$ is done in complete analogy to the one 
for $\ebare \to \pbarp$ \cite{Haidenbauer:2014}. However, unlike the 
situation for $\NbarN$, there is no well established $\lbarl$ interaction potential
available that can be used for the treatment of FSI effects. As a matter of fact, there 
is no direct information about the $\lbarl$ force at all. 
The only constraints one has for that interaction come from studies of FSI effects 
performed for another reaction, namely $\pbarp \to \lbarl$. 
This particular $\pbarp$ annihilation channel has been thoroughly investigated
in the PS185 experiment at LEAR and data are available  
(down to energies very close to the reaction threshold)
for total and differential cross-sections but, thanks to the aforementioned
self-analyzing weak $\Lambda$ decay, also for spin-dependent
observables \cite{PS1852,Barnes:2000,PS185,PS1853}.  
In the present investigation we will resort to $\lbarl$ models that
have been developed for the analysis of those PS185 data by us in the past
\cite{Haidenbauer:1991,Haidenbauer:1992A,Haidenbauer:1992,Haidenbauer:1993}.
Clearly, that introduces unavoidably a model dependence into our $\ebare \to \lbarl$ 
results. But at the present stage we rather view that as an advantage than a drawback. 
Considering various $\lbarl$ models, as we do here, allows us to shed light on the 
issue,
which of the $\ebare \to \lbarl$ observables and, accordingly, which
properties of the $\Lambda$ EMFFs are only weakly influenced by 
detailed aspects of the $\lbarl$ interaction and, thus, can be established
as being practically model-independent, and which quantities show a more pronounced 
sensitivity to variations of the $\lbarl$ interaction and, therefore, could provide
useful information for pinning down the $\lbarl$ force more reliably in the future. 

The paper is organized as follows: 
In Sect.~\ref{sec:form} we describe briefly the employed formalism and in Sect.~\ref{sec:model}
we summarize the properties of the $\lbarl$ potential models used in the
calculations.
Numerical results for the $\ebare \to \lbarl$ reaction are presented in
Sect.~\ref{sec:results}. We show results for observables such as total and differential
cross section but also for spin-dependent quantities like the polarization
and the spin-correlation parameters. In addition, our predictions for the
electromagnetic form factors $G_M$ and $G_E$ are presented.
The paper closes with a summary.

\section{Formalism}
\label{sec:form}

Our formalism for the reaction $\ebare \to \lbarl$ is identical to the 
one developed and described in detail in Ref.~\cite{Haidenbauer:2014} for 
the $\ebare \to \pbarp$ case.
Therefore, we will be very brief here and define only the main quantities. 
We adopt the standard conventions so that the differential cross section
for the reaction $\ebare \to \lbarl$ is given by \cite{Denig}
\be
\frac{d\sigma}{d\Omega} = \frac{\alpha^2\beta}{4 s}
\left[\left| G_M(s) \right|^2 (1+{\rm cos}^2\theta) +
\frac{4M_\Lambda^2}{s} \left| G_E(s) \right|^2 {\rm sin}^2\theta \right]~{\rm .}
\label{eq:diff}
\ee
Here, $\alpha = 1/137.036$ is the fine-structure constant and
$\beta=k_\Lambda/k_e$ a phase-space factor, where $k_\Lambda$ and $k_e$ are the
center-of-mass three-momenta in the $\lbarl$ and $\ebare$ systems, respectively,
related to the total energy via 
$\sqrt{s} = 2\sqrt{M_\Lambda^2+k_\Lambda^2} = 2\sqrt{m_e^2+k_e^2}$.
Further, $M_\Lambda \, (m_e)$ is the $\Lambda$ (electron) mass.
$G_M$ and $G_E$ are the magnetic and electric form factors, respectively.
The cross section as written in Eq.~({\ref{eq:diff}) results from the one-photon
exchange approximation and by setting the electron mass $m_e$ to zero
(in that case $\beta = 2k_\Lambda/\sqrt{s}$). We will restrict
ourselves throughout this work to the one-photon exchange so that the total
angular momentum is fixed to $J=1$ and the $\ebare$ and $\lbarl$ system can be only in
the partial waves $^3S_1$ and $^3D_1$. We use the standard spectral notation
$^{(2S+1)}L_J$, where $S$ is the total spin and $L$ the orbital angular momentum.
The tensor coupling between these two states is taken into account in our calculation.

The integrated reaction cross section is readily found to be
\be
\sigma_{\ebare \to\, \lbarl} = \frac{4 \pi \alpha^2 \beta}{3s}~
\left [ \left| G_M(s) \right|^2 + \frac{2M_\Lambda^2}{s} \left| G_E(s) \right|^2 \right ]~{\rm .}
\label{eq:tot}
\ee

Another quantity used in various analyses is the $\Lambda$ effective form factor $G_{\rm eff}$
which is defined by
\be
|G_{\rm eff} (s)|=\sqrt{\sigma_{\ebare \to\,\lbarl} (s)\over {4\pi\alpha^2
\beta \over 3s} ~ \left [1 +\frac{2M_\Lambda^2}{s}\right ]} \ .
\label{eq:Geff}
\ee

The spin-dependent observables such as the polarization $P$ and the spin-correlation 
parameters $C_{ij}$ are calculated within the formalism used in our study of $\pbarp \to \lbarl$.
Explicit expressions can be found in the Appendix of Ref.~\cite{Haidenbauer:1992} based
on the general parameterization of the spin-scattering matrix for spin-$1/2$ particles 
\cite{Bystricky}.
Expressions for these spin-dependent observables in terms of the form factors $G_M$ and $G_E$ 
can be found in Ref.~\cite{Faldt:2016}, see also \cite{Tomasi:2005,Buttimore:2007}.

\section{The $\lbarl$ interaction}
\label{sec:model}

As already said in the Introduction, 
the $\NbarN$ interaction needed for studies of FSI effects on the timelike 
electromagnetic form factor of the proton is fairly well established. There is a wealth of
empirical information on the elastic ($\pbarp\to\pbarp$) and charge-exchange
($\pbarp\to\nbarn$) reactions from direct scattering experiments and there is 
also a partial-wave analysis available \cite{Zhou:2012}. Thus, for our own investigation 
we could build on a $\NbarN$ potential derived within chiral effective field 
theory, fitted to the results of the PWA \cite{Kang:2013}. 
The situation for the $\lbarl$ is very different, specifically
in this case scattering experiments are impossible.  
Indeed, the only constraints we have for the $\lbarl$ force come from the
FSI in the reaction $\pbarp \to \lbarl$. 
This reaction has been thoroughly studies in the PS185 experiment and 
data are available for total and differential cross-sections but also for 
polarization ($P$) and spin-correlation parameters ($C_{ij}$)
\cite{PS1852,Barnes:2000,PS1853}, down to energies 
very close to the reaction threshold, see also the review in Ref.~\cite{PS185}. 

The reaction $\pbarp \rightarrow \lbarl$ has been analysed in various
model studies, where the strangeness production process is described either 
in terms of the constituent quark model \cite{Kohno,Alberg,Ortega:2011} or 
by the exchange of strange mesons~\cite{Haidenbauer:1991,Haidenbauer:1992}.
In the J\"ulich meson-exchange model the hyperon-production reaction is 
considered within a coupled-channel approach. This allows one to take into 
account rigorously the effects of the initial ($\pbarp$) and final ($\lbarl$) 
state interactions.
The microscopic strangeness production process and the elastic parts of 
the interactions in the $\pbarp$ and $\lbarl$ systems are described by 
meson exchanges, while annihilation processes are accounted 
for by phenomenological optical potentials. To be specific, 
the elastic parts of the $\pbarp$ and $\lbarl$ interactions
are $G$-parity transforms of an one-boson-exchange variant of the
Bonn $NN$ potential~\cite{obepf} and of the hyperon-nucleon model~A of 
Ref.~\cite{Holzenkamp89}, respectively. 
The parameters of the $\pbarp$ annihilation potential are fixed
by an independent fit to $\pbarp$ data in the energy region relevant 
for $\lbarl$ production while those for $\lbarl$ are determined by
a fit to the $\pbarp \to \lbarl$ observables. As documented in 
Refs.~\cite{Haidenbauer:1991,Haidenbauer:1992}, the model achieved a 
fairly good overall description of the PS185 data. 

In the present study we use the $\lbarl$ potentials I, II, and III of 
Ref.~\cite{Haidenbauer:1991} (cf. Table III) and K and QG from 
Ref.~\cite{Haidenbauer:1992} (Table II). The models differ by variations 
in the employed parameterization of the $\NbarN$ annihilation potential
and by differences in the $\pbarp \rightarrow \lbarl$ transition mechanism.
Total cross sections for $\pbarp \to \lbarl$ of the considered potentials 
agree with each other and with the experiment up to 
$p_{\rm lab} \approx 1700$~MeV (corresponding to $\sqrt{s}\approx 2.32$~GeV  
or an excess energy $Q=\sqrt{s}-2m_\Lambda$ of about $90$~MeV).
However, the spin-dependent observables are not always reproduced quantitatively.
Still we believe that it is useful to employ all models in the present study 
because this allows us to shed light on the (unavoidable) model dependence of 
our predictions for the electromagnetic form factor of the $\Lambda$ in the timelike region. 

\begin{figure}[t]
\begin{center}
\includegraphics[height=85mm,angle=-90]{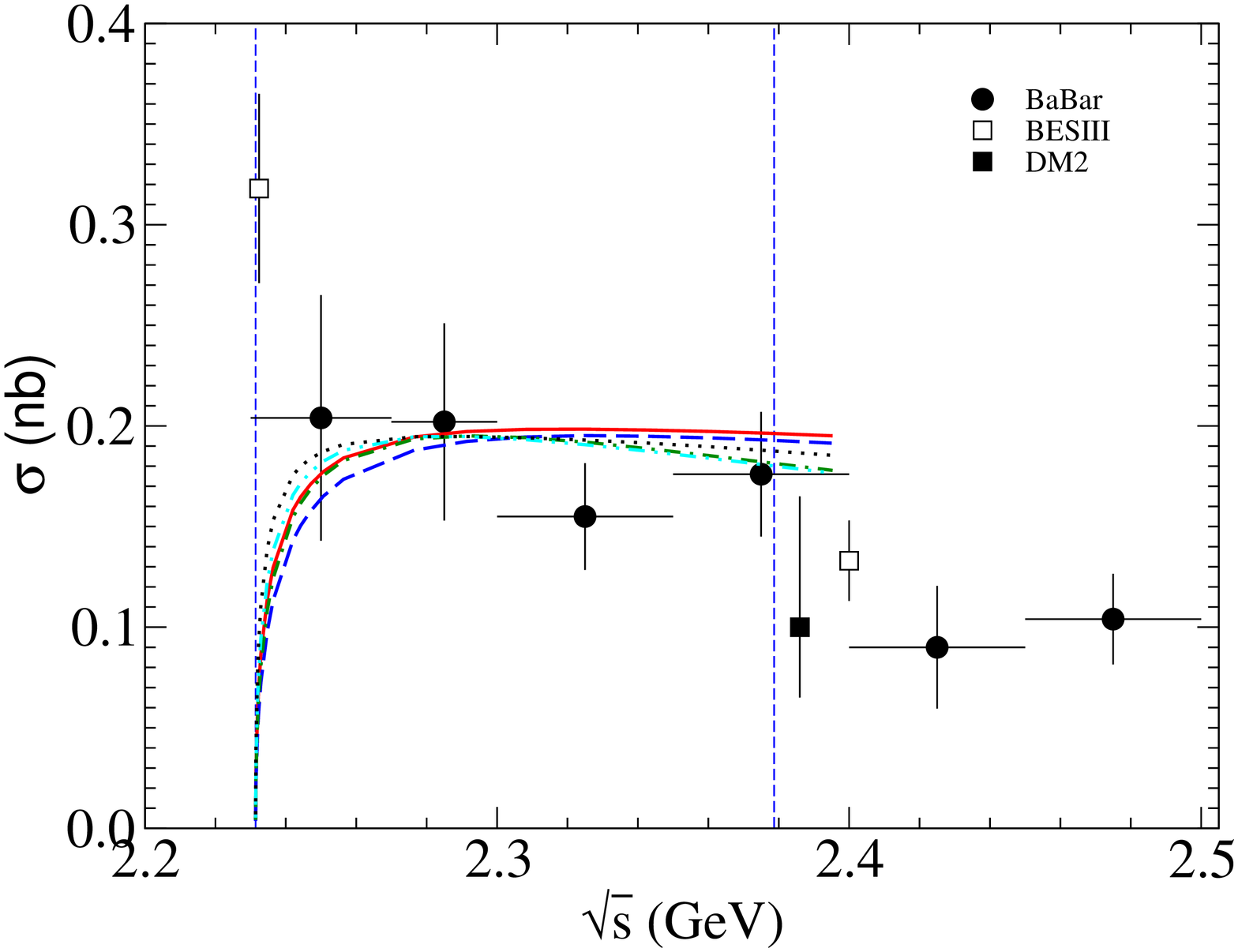}\includegraphics[height=85mm,angle=-90]{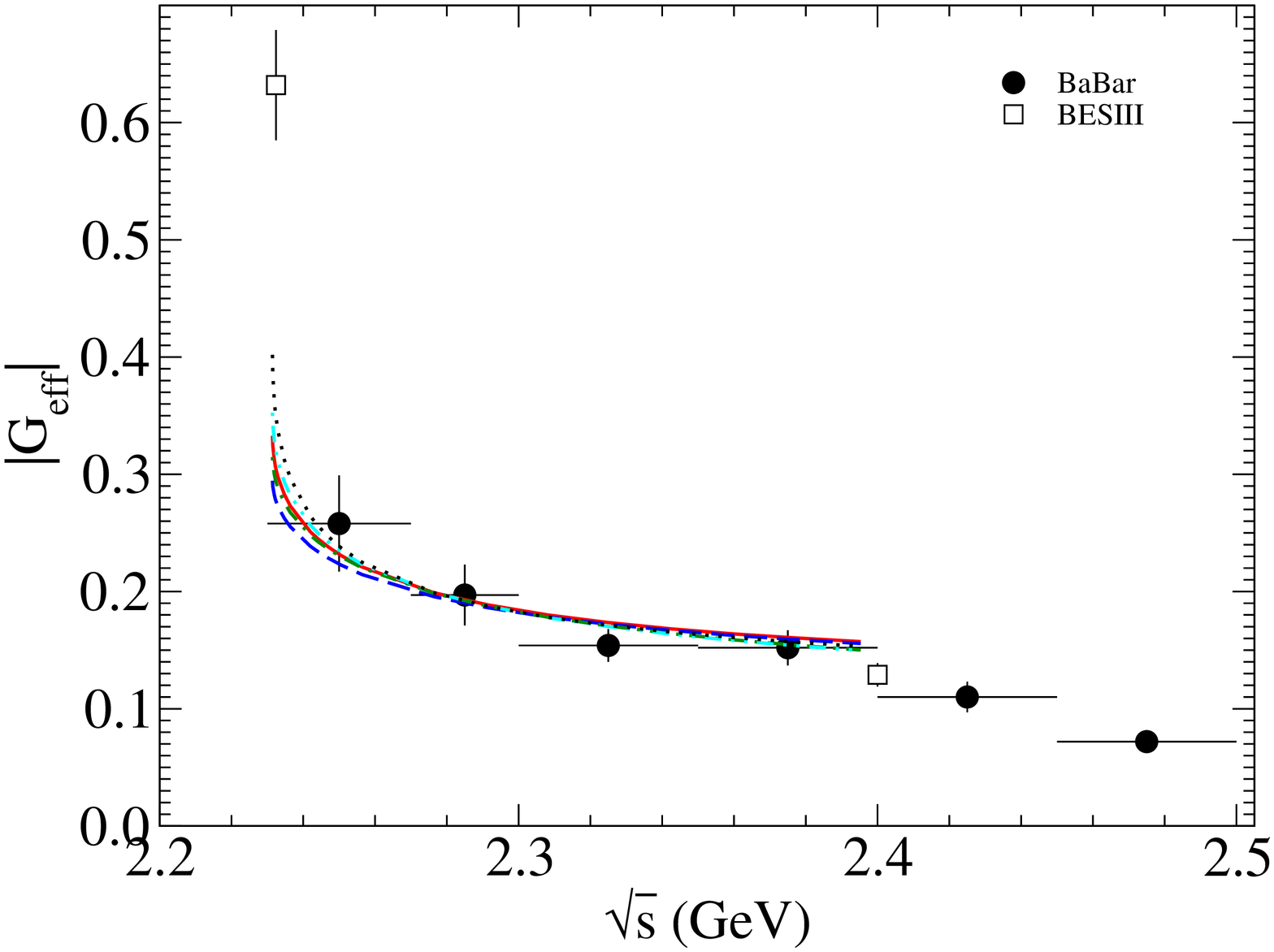}
\caption{Total cross section (left) and effective form factor $|G_{\rm eff}|$ (right)
for $\ebare \rightarrow \lbarl$.
Data are from the DM2 \cite{Bisello:1990}, BaBar \cite{Aubert:2007}, and
BESIII \cite{Morales:2016} collaborations. Note that the latter data are still preliminary.
The vertical lines indicate the $\lbarl$ and $\spbarsp$ thresholds,
respectively.
The solid, dashed, and dash-dotted lines correspond to the $\lbarl$ models I, II, and II
from Ref.~\cite{Haidenbauer:1991}, the dash-double-dotted and dotted lines
to the models K and Q described in Ref.~\cite{Haidenbauer:1992}.
}
\label{fig:tot}
\end{center}
\end{figure}

\section{Results and discussion}
\label{sec:results}

Results for the $\ebare \to \lbarl$ cross section are depicted in Fig.~\ref{fig:tot} 
(left panel). The $\lbarl$ threshold is at $\sqrt{s}=2.23138$~GeV in our calculation.
We normalized the various curves to yield the same value at 
the maximum (i.e. roughly $0.2\,$nb) so that one can easily compare the differences
in the energy dependence. The overall normalization, corresponding to the value
of the bare electromagnetic form factors $G^0_M$ and $G^0_E$, is actually the only
free parameter in our calculation, see Ref.~\cite{Haidenbauer:2014}. 
Obviously, for all $\lbarl$ models the cross section rises 
rather sharply from the threshold and then remains practically constant for the next 
$100$~MeV or so. This behavior is very well in line with the experimental information 
from the BaBar collaboration \cite{Aubert:2007}. There are quantitative differences
between the predictions of the different $\lbarl$ potentials considered but it 
is reassuring to see that, overall, the model dependence is fairly moderate. 
 
Thus, it seems that despite of all the uncertainties in the dynamics reflected in the
various considered $\lbarl$ models, the data on $\pbarp \to \lbarl$ put fairly tight 
constraints on the properties of the $\lbarl$ interaction. 
However, equally or possibly even more important is presumably the fact that 
the $\lbarl$ FSI employed here incorporates some very essential features.
First, it is generated by solving a scattering equation and, therefore, 
properly unitarized, and secondly it includes effects from the presence of 
annihilation channels. We believe that most likely those features alone fix already
the qualitative behavior of the $\ebare \to \lbarl$ near-threshold cross section.
This conjecture is supported by the situation in the $\ebare \to \pbarp$ reaction.
In this case the FSI is very different when it comes to details. For example, in
the $\NbarN$ system there is a contribution from the important and long-ranged 
pion-exchange that creates a strong tensor force. In case of $\lbarl$ one pion 
exchange cannot contribute because of isospin symmetry. Still the near-threshold behaviour 
of the $\ebare \to \lbarl$ and $\ebare \to \pbarp$ cross sections is rather similar,
compare Fig.~\ref{fig:tot} with Fig.~3 in Ref.~\cite{Haidenbauer:2014}
(cf. also Fig.~2 in \cite{Haidenbauer:2015} for $\ebare \to \nbarn$). 
It is remarkable that even the presence of the Coulomb interaction in the $\pbarp$ case, 
usually accounted for in terms of the Sommerfeld-Gamov factor \cite{Haidenbauer:2014}, 
has very little impact on the qualitative similarity, 
despite the fact that it changes the threshold behavior of the $\ebare \to \pbarp$
reaction so that the cross section remains finite even at the nominal threshold.

\begin{figure}[htb!]
\begin{center}
\includegraphics[height=85mm,angle=-90]{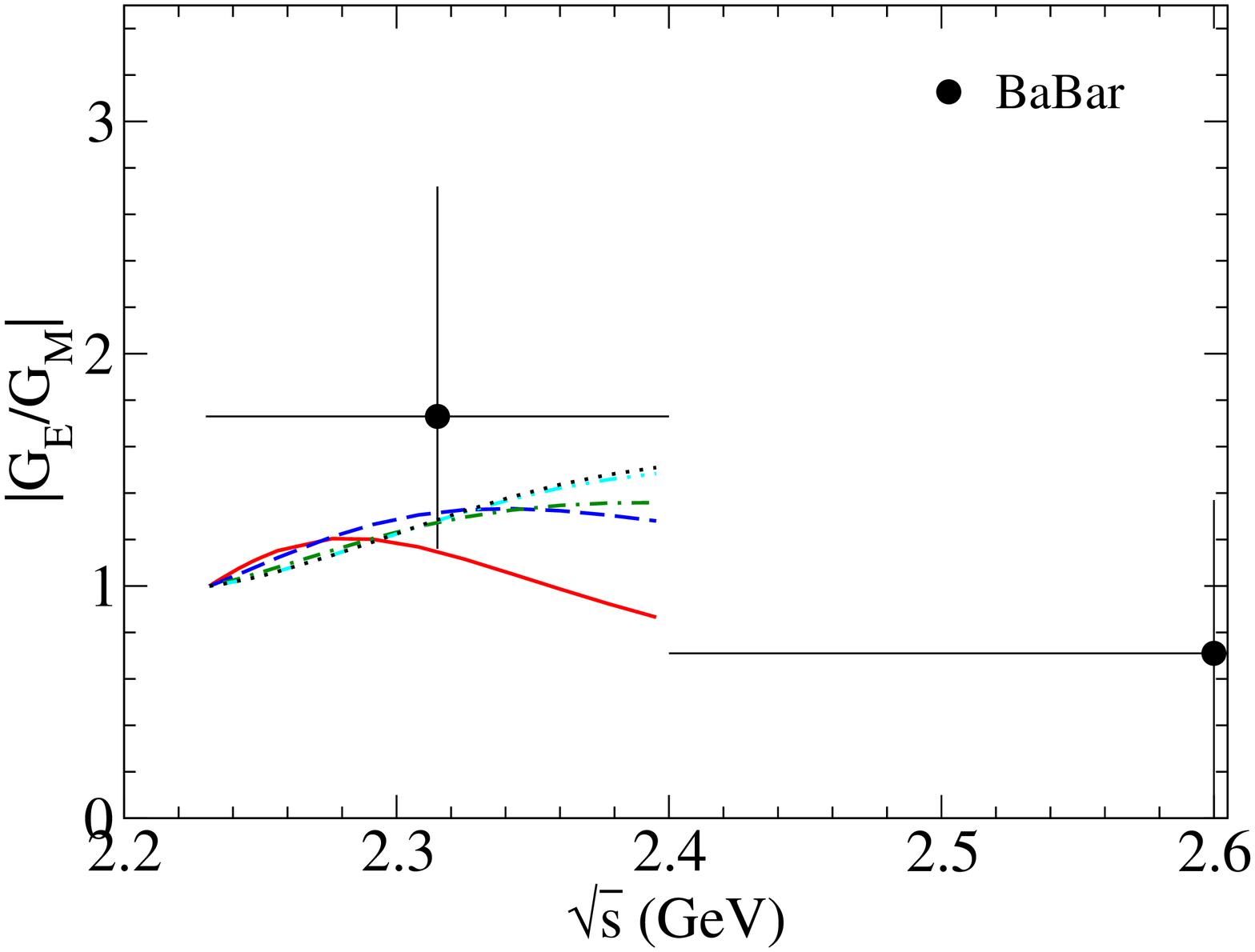}\includegraphics[height=85mm,angle=-90]{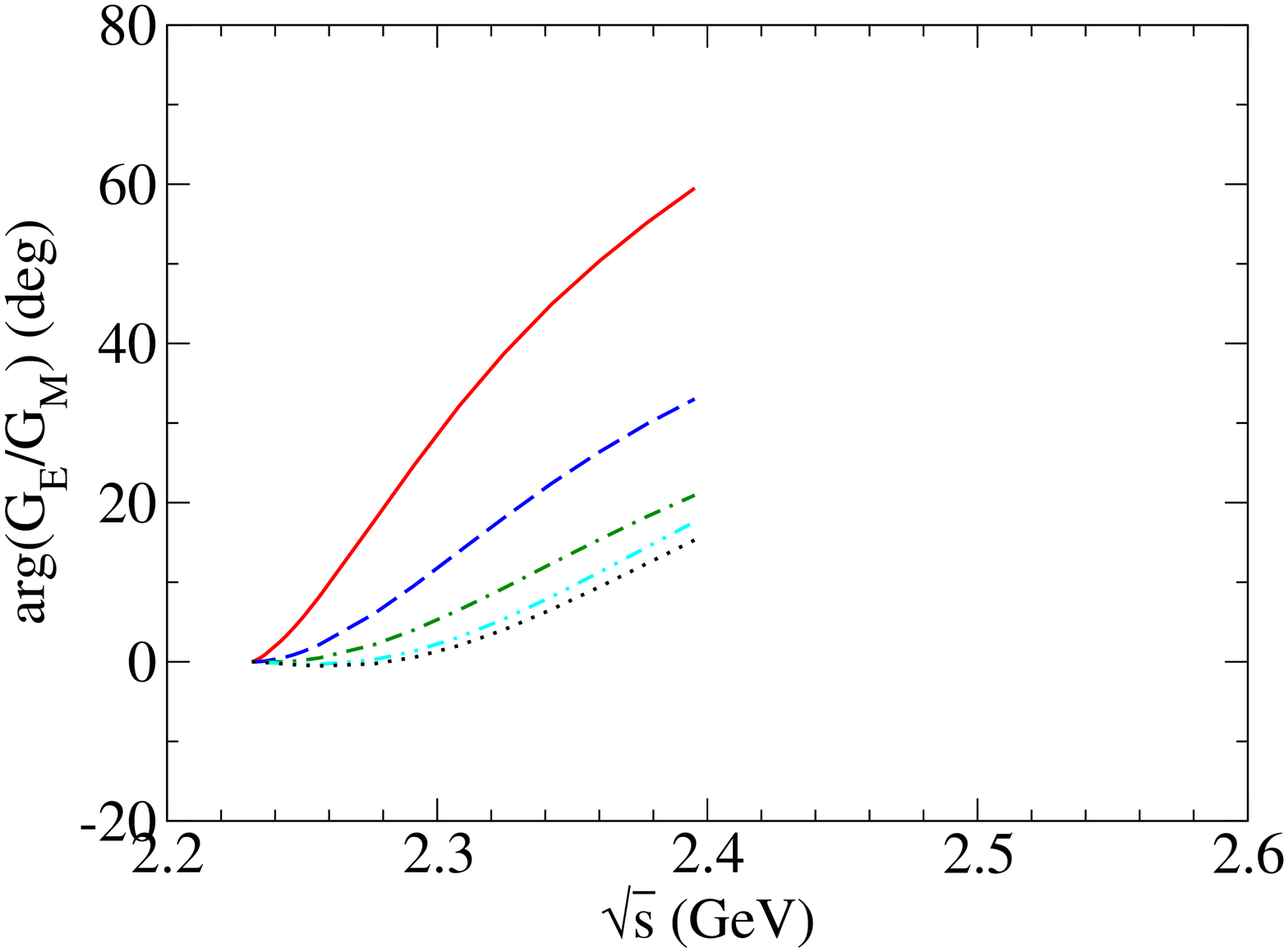}
\caption{The ratio $|G_E/G_M|$ (left) and phase $\phi=\rm{arg}(G_E/G_M)$ (right) as a function of the 
total cms energy. Data are from Ref.~\cite{Aubert:2007}.
For notation, see Fig.~\ref{fig:tot}.
}
\label{fig:GEGM}
\end{center}
\end{figure}

\begin{figure}[htb!]
\begin{center}
\includegraphics[height=85mm,angle=-90]{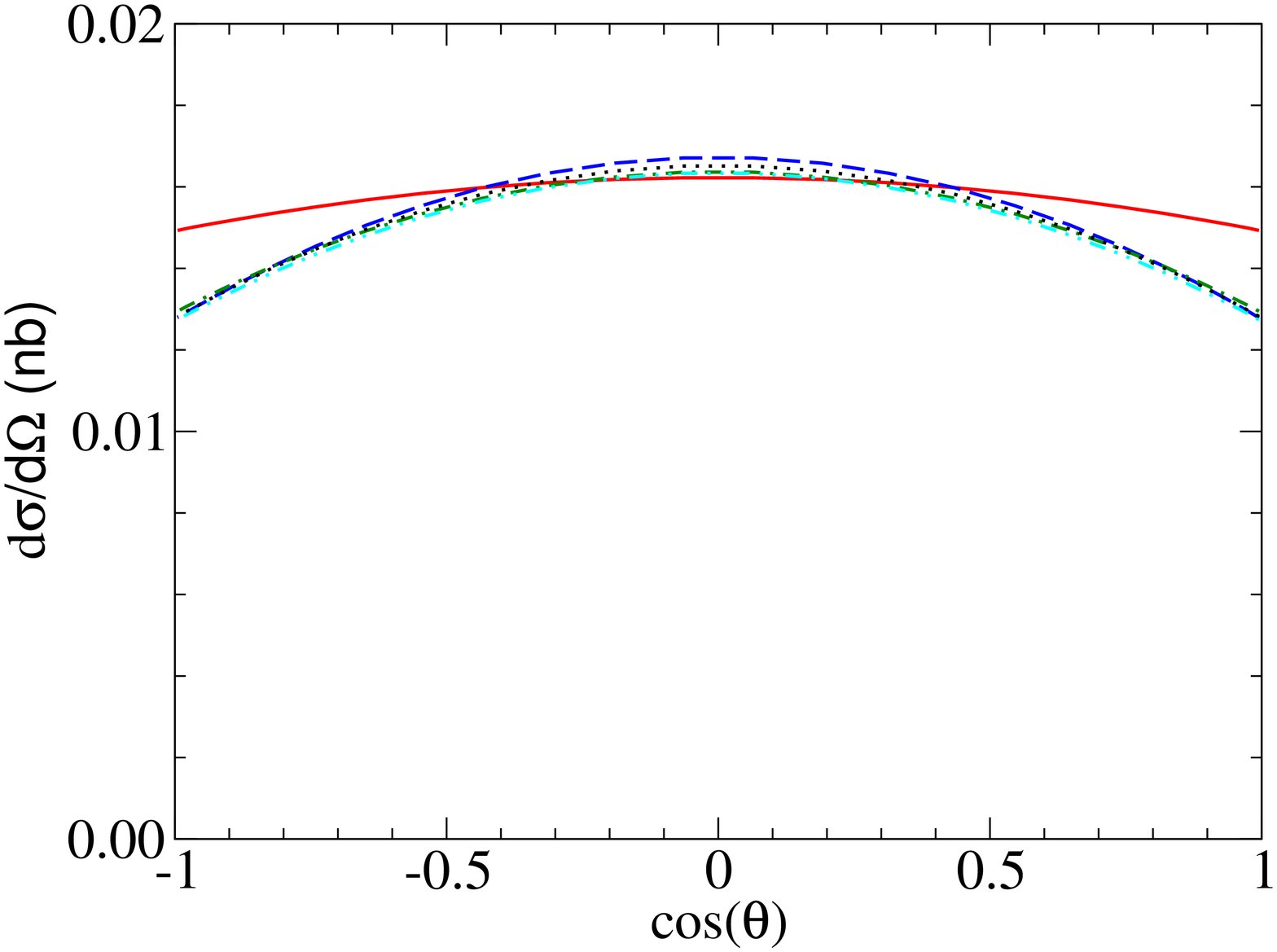}\includegraphics[height=85mm,angle=-90]{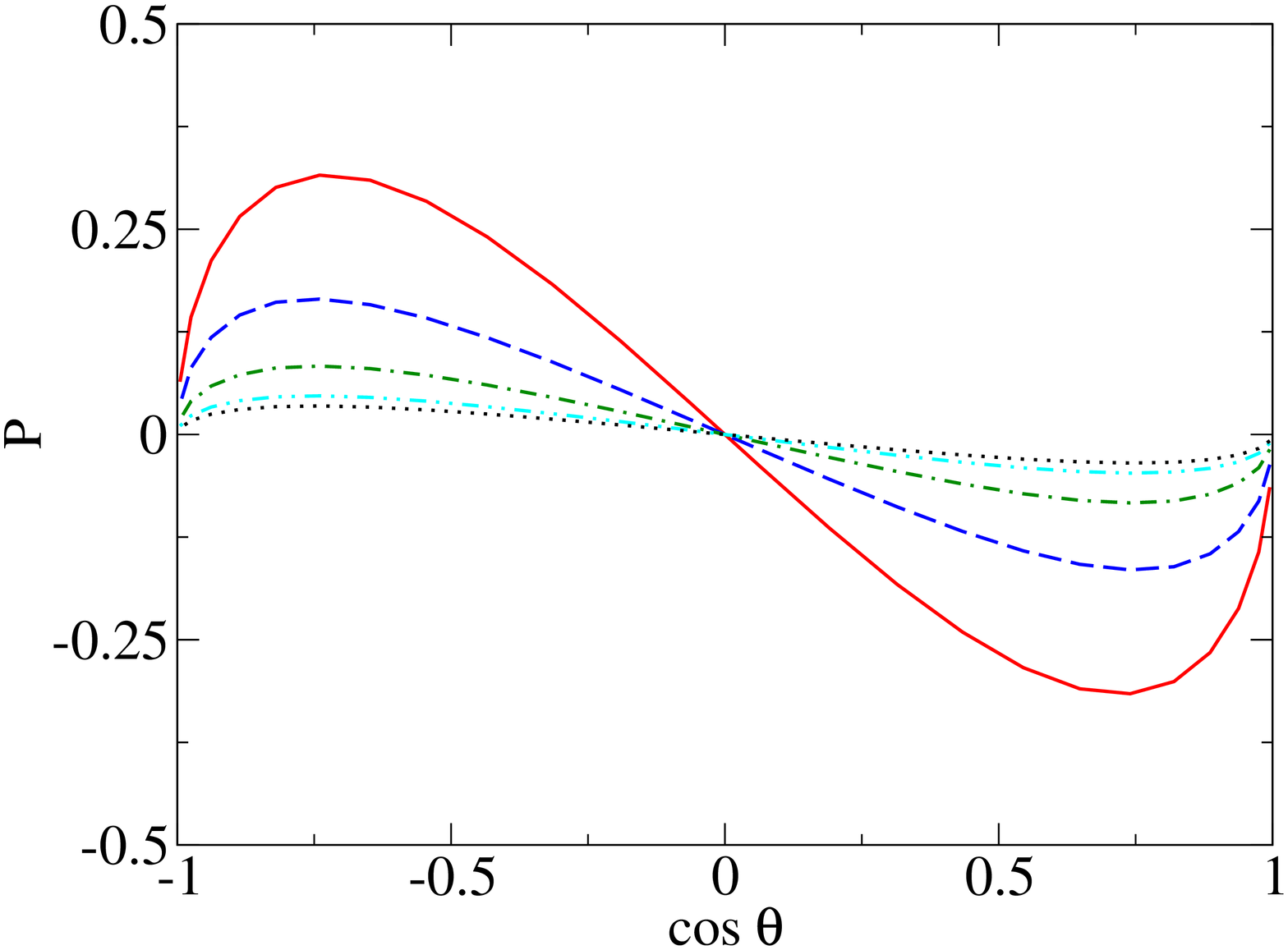}
\caption{Differential cross section (left) and polarization (right) for $\ebare \to \lbarl$ 
at the excess energy $Q=90$~MeV ($\sqrt{s}=2.32$~GeV).
For notation, see Fig.~\ref{fig:tot}.
}
\label{fig:Diff2}
\end{center}
\end{figure}

\begin{figure}[htb!]
\begin{center}
\includegraphics[height=165mm,angle=-90]{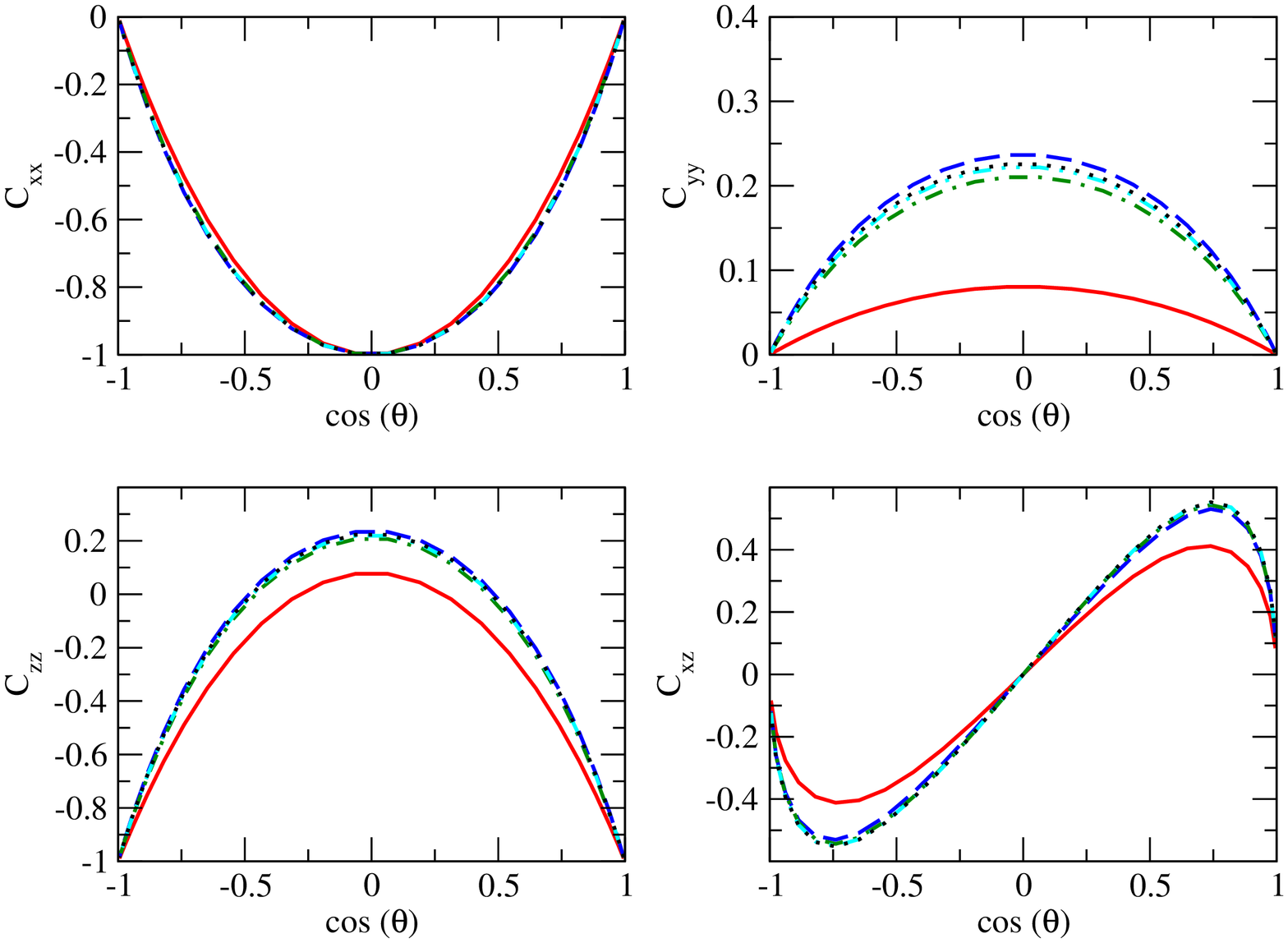}
\caption{Spin correlations parameters for $\ebare \to \lbarl$ 
at the excess energy $Q=90$~MeV ($\sqrt{s}=2.32$~GeV). 
For notation, see Fig.~\ref{fig:tot}.
}
\label{fig:Spin2}
\end{center}
\end{figure}

Fig.~\ref{fig:tot} contains also two new and still preliminary data points from the
BESIII collaboration \cite{Morales:2016} (open squares). Those data suggest a very 
different trend for the energy dependence of $\sigma_{\ebare \to \lbarl}$. Specifically,
a large finite value for the cross section practically at the threshold is suggested.
This cannot be reproduced by our calculation because the phase-space factor $\beta$
(see Eq.~(\ref{eq:tot})) enforces the cross section to go to zero linearly with the
cms momentum $k_\Lambda$ of the $\lbarl$ system. There is no Coulomb interaction here 
that would change the threshold behavior. The only possibility could be a very narrow 
resonance sitting more or less directly at the threshold which would then allow to 
overrule the $k_\Lambda$ behavior from the phase space alone.
Including such a threshold resonance in our $\lbarl$ potentials
would, however, completely spoil the agreement with the $\pbarp \to \lbarl$ experiments. 
Anyway, one has to wait simply on the final results of the BESIII measurement.
Interestingly, initial data for $\pbarp \to \lbarl$ suggested that there could 
be a near threshold resonance in the $^3S_1$ $\lbarl$ state \cite{Carbonell:1993}.
However, a later high-statistics measurement by Barnes et al.~\cite{Barnes:2000}
ruled that out convincingly. This experiment scrutinized the first $5$ MeV above 
the $\lbarl$ threshold with an unprecedented resolution of $0.2$ MeV.

For illustration purposes we show results up to the $\spbarsp$ threshold
(indicated by a vertical line). However, we expect that near that threshold
$\ebare \to \sbars \to \lbarl$ two-step processes will become important, 
which are not included, and, therefore, our calculation is certainly no longer
valid at those energies. In this context it is interesting to note that, 
indeed, the data suggest a certain drop in the $\ebare \to \lbarl$ cross section 
right above the $\spbarsp$ threshold, cf. Fig.~\ref{fig:tot} (left).  
In reality we would estimate the validity range of the predictions
to about 100~MeV from the threshold, i.e. up to $\sqrt{s}\approx 2.32$~GeV. This 
is the energy range where the $\pbarp \to \lbarl$ data are described by the potentials 
and it is also the range where we know from the $\ebare \to \pbarp$ case that FSI 
effects are relevant and determine the energy dependence of the observables \cite{Haidenbauer:2014}. 
Over a larger energy region, the intrinsic energy- and momentum dependence of the
$\lbarl$ production mechanism itself should become more significant or even dominant 
and then the present assumption that the bare electromagnetic form factors $G^0_M$ 
and $G^0_E$ that enter our calculation,
see Eq.~(10) of Ref.~\cite{Haidenbauer:2014} for more details, 
are constant is no longer valid.  

Predictions for the $\Lambda$ effective form factor $G_{\rm eff}$ 
are presented in the right panel of Fig.~\ref{fig:tot}. Since that quantity differs from 
the cross section only by kinematical factors, see Eq.~(\ref{eq:Geff}), 
it is not surprising that our results are again in line with the corresponding 
empirical information deduced from the experiments cross sections \cite{Aubert:2007}. 
Once again one can see that the preliminary near-threshold data point from BESIII differs 
drastically from the general trend. 

Results for the form factor ratio $|G_E/G_M|$ and the relative phase between $G_E$ and $G_M$
are presented in Fig.~\ref{fig:GEGM}. Clearly, those quantities are much more model-dependent
and specifically for the phase there are fairly large variations between the predictions from
the $\lbarl$ potentials considered. The experimental results for $|G_E/G_M|$ from 
Ref.~\cite{Aubert:2007} are also shown. The lower mass resolution of that quantity 
does not allow any more concrete conclusions. However, the trend that the ratio is 
somewhat larger than 1 for smaller
energies and possibly smaller than 1 for larger energies is also suggested by most of the
$\lbarl$ potentials. 
Let us also mention that the relative phase between $G_E$ and $G_M$ was determined to be
in the range $-0.76 < \sin \phi < 0.98$ by the BaBar collaboration \cite{Aubert:2007}. 

Finally, and in anticipation of pertinent results from the BESIII collaboration \cite{Morales:2016,Tord},
in Figs.~\ref{fig:Diff2} and \ref{fig:Spin2} we present exemplary predictions for the differential cross 
section and for polarization observables at an excess energy of $90$~MeV. As discussed above,
this energy might be alrighty close to the limit of validity of our calculation. 
However, at higher energies there is a stronger model dependence and we believe that it 
is instructive to see how the resulting differences, already discussed in the context of the
form factors $G_E$ and $G_M$, manifest themselves directly in the actual observables accessible 
in the experiment. 
A view on the sensitivity of specific observables to variations of the $\lbarl$ interaction, or
equivalently of $G_M$ and $G_E$, could be quite helpful for planning future experiments. 

Among the observables presented in Figs.~\ref{fig:Diff2} and \ref{fig:Spin2} the polarization
$P$ exhibits the strongest variation with the employed $\lbarl$ interaction. This is not
surprising because $P$ is proportional to ${\rm Im} (G_M\,G^*_E)$, see \cite{Faldt:2016},
i.e. it depends strongly on the relative phase between $G_M$ and $G_E$, and we have seen
above that there are sizable differences in the prediction for this quantity. 
The spin-correlation parameter $C_{xz}$ is proportional to ${\rm Re} (G_M\,G^*_E)$ 
\cite{Faldt:2016} and, therefore, reflects mainly variations in the ratio $|G_E/G_M|$. 
These variations can be seen also, and more prominently, in $C_{yy}$ which is sensitive 
to the difference $|G_M|^2 - |G_E|^2$. 

Note that the sign of some polarization observables depends on the choice of the reference 
frame. As already said in Sec. II, we adopt here the formalism described in the Appendix of 
Ref.~\cite{Haidenbauer:1992}. The spin-observables are calculated in the $\lbarl$ 
coordinate system, i.e. the direction of the $\bar \Lambda$ is defined as $z$ direction.
This agrees with the one employed in the PS185 experiment in the analysis of the 
$\pbarp \to \lbarl$ reaction \cite{PS1852}.  
The spin observables in Ref.~\cite{Faldt:2016} have partly opposite signs, in 
particular $A_y=-P$, $A_{xx}=-C_{xx}$, and $A_{zz}=-C_{zz}$.

As already noted above, we believe that the similar properties of the near-threshold
cross sections for $\ebare \to \lbarl$ and $\ebare \to \pbarp$, consisting in a sharp
rise and then a practically constant behavior,
can be understood in terms of FSI effects driven by qualitative aspects that are 
common to the $\lbarl$ and $\pbarp$ interactions, like unitarity and the presence
of annihilation processes. 
Available data for $\pbarp\to\sbarl$ and $\pbarp\to\sobarso$ from the BaBar 
collaboration \cite{Aubert:2007} suggest likewise such a behavior.
However, since the mass resolution of the data for those reactions is 
significantly lower one cannot draw firm conclusions at the moment.
There are data with comparable resolution (i.e. $20$~MeV/c$^2$) for the reaction 
$\ebare \to \lcbarlc$ from the Belle collaboration \cite{Pakhlova:2008}.
Those data were analysed by us in Ref.~\cite{Guo:2010} and it was found that
the inclusion of FSI effects (based on a $\lcbarlc$ interaction derived 
from the $\lbarl$ interactions employed in the present study by 
invoking SU(4) flavor-symmetry arguments \cite{Haidenbauer:2009})
improves the description of the data in the threshold region.
On the other hand, in this reaction the near-threshold invariant mass spectrum 
is dominated by the so-called $X$(4630) resonance and, therefore, exhibits a behavior 
that clearly differs from the ones in the $\pbarp$ and $\lbarl$ channels. 

\section{Summary}

In the present paper we investigated the reaction $\ebare \to \lbarl$
in the near-threshold region with specific emphasis on the role played
by the interaction in the final $\lbarl$ state. The
calculation is based on the one-photon approximation for the elementary
reaction mechanism, but takes into account rigorously the effects of
the $\lbarl$ interaction in close analogy to our work on $\ebare \to \pbarp$
\cite{Haidenbauer:2014}. 
For the $\lbarl$ interaction we utilized a variety of potentials 
that were constructed for the analysis of the reaction $\pbarp \to \lbarl$ 
about two decades ago \cite{Haidenbauer:1991,Haidenbauer:1992}. Those
potentials are basically of phenomenological nature but fitted and 
constrained by the wealth of near-threshold data on $\pbarp \to \lbarl$
taken in the PS185 experiment at LEAR \cite{PS185}.  

The energy dependence of the near-threshold $\ebare \to \lbarl$ cross section 
reported by the BaBar collaboration \cite{Aubert:2007}, which consists in a sharp 
rise from the threshold and then a flat behavior for the next $100$~MeV or so,
is well reproduced by our calculation based on the various $\lbarl$ potentials.
Since we see only a moderate model dependence in our results we believe that, 
most likely, general features like that the employed FSI is generated by solving a 
scattering equation and, therefore, properly unitarized, and that it includes 
effects from the presence of annilation channels  
alone fix already the qualitative behavior of the $\ebare \to \lbarl$ 
near-threshold cross section.
Indeed the situation is very similar to that in the $\ebare \to \pbarp$ reaction 
where the measureed near-threshold cross section shows a comparable
behavior and where again FSI effects due to the $\pbarp$ interaction are able
to account for this \cite{Haidenbauer:2014}. 

Prelimary results from the BESIII collaboration indicate a very much different
energy dependence of the $\ebare \to \lbarl$ cross section \cite{Morales:2016}.
If that behavior is confirmed in the final analysis it will be very difficult to 
reconcile the $\ebare$ results with our knowledge on the $\pbarp \to \lbarl$
reaction \cite{PS185}. Note that the latter is also completely dominated by the $^3S_1$ 
partial wave close to threshold \cite{PS1852,Haidenbauer:1991} and thus by exactly 
the same $\lbarl$ FSI. 
In any case, further data in the near-threshold region with better mass
resolution would be very helpful to resolve this issue.

Finally, 
due to the self-analyzing character of the weak $\Lambda$ decay  
the polarization as well as spin-correlation parameters for the reaction
$\ebare \to \lbarl$ can be determined, in close analogy to what has been already 
achieved in the PS185 experiment for $\pbarp \to \lbarl$ \cite{PS185}.
We are looking forward to analyze such data within our formalism, once they
become available, because they will certainly allow one to put much tighter 
constraints on the interaction in the $\lbarl$ system. 

\vskip 0.2cm 
\noindent
{\bf Acknowledgments}
We acknowledge stimulating discussions with Tord Johansson. 
This work is supported in part by the DFG and the NSFC through
funds provided to the Sino-German CRC 110 ``Symmetries and
the Emergence of Structure in QCD''. The work of UGM was also 
supported by the Chinese Academy of Sciences (CAS) President's
International Fellowship Initiative (PIFI) (Grant No. 2015VMA076).

%

\end{document}